\documentclass{amsart}

\usepackage{fancyhdr}
\usepackage{lastpage}
\usepackage{stmaryrd,yhmath}

\newcommand{\bdis}{\begin{displaymath}}
\newcommand{\edis}{\end{displaymath}}
\newcommand{\be}{\begin{equation}}
\newcommand{\ee}{\end{equation}}

\newcommand{\mcal}{\mathcal}

\newcommand{\zf}{\zeta\left(\frac{1}{2}+it\right)}

\newtheorem{lemma}[]{Lemma}

\theoremstyle{definition}

\newtheorem{cor}[]{Corollary}

\theoremstyle{remark}
\newtheorem{remark}[]{Remark}

\newtheorem*{mydef1}{{\bf Theorem}}

\newtheorem*{mydef61}{{\bf Formula 1}}

\newtheorem*{mydef62}{{\bf Formula 2}}

\newtheorem*{mydef7}{{\bf Definition}}

\numberwithin{equation}{section}

\begin{document}

\title{Riemann's hypothesis and some infinite set of microscopic universes of the Einstein's type in the early period of the evolution of
the Universe}

\author{Jan Moser}

\address{Department of Mathematical Analysis and Numerical Mathematics, Comenius University, Mlynska Dolina M105, 842 48 Bratislava, SLOVAKIA}

\email{jan.mozer@fmph.uniba.sk}

\keywords{Riemann zeta-function}

\begin{abstract}
We obtain in this paper, as a consequence of the Riemann hypothesis, certain class of topological deformations of the graph of the function
$|\zf|$. These are used to construct an infinite set of microscopic universes (on the Planck's scale) of the Einstein type. \\
Dedicated to the 90th anniversary of the A.S. Edington's book \emph{The mathematical theory of relativity}.
\end{abstract}

\maketitle

\section{Introduction and the main Result}

\subsection{}

Let (see \cite{13}, pp. 79, 329)
\be \label{1.1}
\begin{split}
& Z(t)=e^{i\vartheta(t)}\zf , \\
& \vartheta(t)=-\frac t2\ln\pi+\text{Im}\ln\Gamma\left(\frac 14+\frac 12 it\right)= \\
& = \frac t2\ln\frac{t}{2\pi}-\frac t2-\frac{\pi}{8}+\mcal{O}\left(\frac 1t\right).
\end{split}
\ee
We denote by $\{\gamma\}$ the sequence of the roots of the equation
\bdis
Z(t)=0,
\edis
and, further, we denote by $\{t_0\}$ the sequence of the roots of the equation
\bdis
Z'(t)=0,\ t_0\not=\gamma;\ Z(t_0)\not=0.
\edis

\begin{remark}
On the Riemann hypothesis, the points of the sequences $\{\gamma\}$, $\{ t_0\}$ are separated, i. e.
\bdis
\gamma'<t_0<\gamma'',
\edis
where $\gamma',\gamma''$ are neighboring points of the sequence $\{\gamma\}$, (comp. \cite{7}).
\end{remark}

We have proved in our paper \cite{9} the following theorem: on the Riemann hypothesis we have
\be \label{1.2}
\frac{Q(t_0)}{m(t_0)}<t_0\ln^2t_0\ln_2t_0\ln_3t_0,\quad t_0\to\infty,
\ee
where
\bdis
\ln_2t_0=\ln\ln t_0,\dots ,
\edis
and
\be \label{1.3}
Q(t_0)=\max\{ \gamma''-t_0,t_0-\gamma'\},\quad m(t_0)=\min\{ \gamma''-t_0,t_0-\gamma'\}.
\ee

\begin{remark}
The sequence $\{ t_0\}$ oscillates in a complicated manner around the sequence $\{ \gamma\}$. Consequently, the quotient
\bdis
\frac{Q(t_0)}{m(t_0)}
\edis
characterizes the asymmetry of the point $t_0$ relatively to the points $\gamma',\gamma''$.
\end{remark}

The estimate (\ref{1.2}) follows from the formula
\bdis
\frac{\pi}{4}\sim\sum_{\gamma}\frac{t_0}{\gamma^2-t_0^2},\quad t_0\to\infty,
\edis
(see \cite{7}) and this is the conjugate formula to the Riemann formula
\bdis
c+2-\ln 4\pi=\sum_{\gamma}\frac{1}{\frac 14+\gamma^2},
\edis
where $c$ is the Euler constant.

\subsection{}

In this paper we will study the following quotient
\bdis
0<\frac{Z(t)}{Z(t_0)},\quad t\in (\gamma',\gamma''),\ \gamma'<t_0<\gamma'',\quad \gamma'\to\infty,
\edis
where $Z(t_0)$ is the local maximum or local minimum of the function $Z(t)$. We make use the estimate (\ref{1.2}) to prove the following theorem.

\begin{mydef1}
If
\be \label{1.4}
\begin{split}
& \Omega(t_0)=t_0\ln^4t_0+\ln t_0\ln w(t_0), \\
& w(t_0)=\max\left\{ \frac{2}{\omega(t_0)}t_0\ln^3t_0,\frac{2}{m(t_0)}\right\},\ \omega(t_0)\in (0,1), \\
& \Delta_1(t_0)=\{ 1-\omega(t_0)\}(t_0-\gamma'),\ \Delta_2(t_0)=\{ 1-\omega(t_0)\}(\gamma''-t_0), \\
& J(t_0)=[t_0-\Delta_1(t_0),t_0+\Delta_2(t_0)],
\end{split}
\ee
then, on the Riemann hypothesis, we have
\be \label{1.5}
e^{-A\Omega(t_0)}<\frac{Z(t)}{Z(t_0)}\leq 1,\quad t\in J(t_0)\subset (\gamma',\gamma''),
\ee
where, of course, the lower estimate in (\ref{1.5}) is the nontrivial result.
\end{mydef1}

\begin{remark}
Since (see (\ref{1.4}))
\bdis
\text{mes}\{ (\gamma',\gamma'')\setminus J(t_0)\}=\omega(t_0)(\gamma''-\gamma'),
\edis
then the interval $J(t_0)$ is the essential part of the interval $(\gamma',\gamma'')$ if $\omega(t_0)$ is sufficiently small, for example $\omega(t_0)=10^{-58},\dots$
\end{remark}

\subsection{}

Now we define the sequence $\{\alpha(t_0)\}$ by the following condition
\be \label{1.6}
\alpha(t_0)=\frac{\omega^4(t_0)m^4(t_0)}{t_0\Omega(t_0)},\quad t_0>K
\ee
for sufficiently big $K>0$. Since the Littlewood estimate (see \cite{5}, p. 237)
\be \label{1.7}
\gamma''-\gamma'<\frac{A}{\ln\ln\gamma'},\quad \gamma'\to\infty
\ee
holds true on the Riemann hypothesis, then by (\ref{1.4}), (\ref{1.7})
\be \label{1.8}
m(t_0),Q(t_0)<\frac{A}{\ln\ln\gamma'},\quad t_0 \to\infty.
\ee
Hence, we have from (\ref{1.6}) by (\ref{1.4}), (\ref{1.8})
\be \label{1.9}
\alpha(t_0)\Omega(t_0)=\frac{1}{t_0}\omega^4(t_0)m^4(t_0)<\frac{B}{t_0(\ln\ln t_0)^2},
\ee
i. e. we obtain from (\ref{1.5}) by (\ref{1.9})
\be \label{1.10}
1\geq \left\{\frac{Z(t)}{Z(t_0)}\right\}^{\alpha(t_0)}>e^{-\alpha(t_0)\Omega(t_0)}=1+\mcal{O}\left(\frac{1}{t_0}\right),
\ee
i. e. we have the following

\begin{cor}
On the Riemann hypothesis the following asymptotic formula
\be \label{1.11}
\left\{\frac{Z(t)}{Z(t_0)}\right\}^{\alpha(t_0)}=1+\mcal{O}\left(\frac{1}{t_0}\right),\quad t\in J(t_0),\ t_0\to\infty
\ee
holds true.
\end{cor}

\section{Formulae for the logarithmic derivatives of the function $Z(t)$ and some lemmas}

\subsection{}

The following main formula follows from the Riemann hypothesis (see \cite{7}, (1))
\be \label{2.1}
-\frac{{\rm d}}{{\rm d}t}\left\{\frac{Z'(t)}{Z(t)}\right\}=\sum_\gamma \frac{1}{(t-\gamma)^2}+\mcal{O}\left(\frac{1}{t}\right),\ t\not=\gamma.
\ee
Since the series in (\ref{2.1}) is uniformly convergent on $J(t_0)$ and $Z'(t_0)=0$, then we obtain by integration of (\ref{2.1}) in the limits $t_0,t\in J(t_0)$  (comp. \cite{8}, (7)) the following

\begin{mydef61}
We have on the Riemann hypothesis
\be \label{2.2}
-\frac{Z'(t)}{Z(t)}=(t-t_0)\sum_\gamma\frac{1}{(t-\gamma)(t_0-\gamma)}+\mcal{O}\left(\frac{|t-t_0|}{t_0}\right),\ t\in J(t_0).
\ee
\end{mydef61}

Since
\bdis
(t-t_0)\sum_\gamma\frac{1}{(t-\gamma)(t_0-\gamma)}=\sum_\gamma\left(\frac{1}{t_0-\gamma}-\frac{1}{t-\gamma}\right)
\edis
then we obtain by integration of (\ref{2.2}) in the limits $t_0,t\in J(t_0)$ the following
\begin{mydef62}
On the Riemann hypothesis
\be \label{2.3}
-\ln\frac{Z(t)}{Z(t_0)}=\sum_\gamma\left\{\frac{t-t_0}{t_0-\gamma}-\ln\left|\frac{t-\gamma}{t_0-\gamma}\right|\right\}+\mcal{O}\left\{\frac{(t-t_0)^2}{t_0}\right\},\ t\in J(t_0).
\ee
\end{mydef62}

\subsection{}

Next, the following lemmas hold true.

\begin{lemma}
\be \label{2.4}
\sum_\gamma\frac{1}{(t-\gamma)^2}=\mcal{O}\left\{\frac{\ln t_0}{\omega^2(t_0)m^2(t_0)}\right\},\quad t\in J(t_0).
\ee
\end{lemma}

\begin{lemma}
On the Riemann hypothesis we have
\be \label{2.5}
\left\{\frac{Z'(t)}{Z(t)}\right\}^2=\mcal{O}\left\{\frac{\ln^2t_0}{\omega^4(t_0)m^4(t_0)(\ln\ln t_0)^2}\right\},\quad t\in J(t_0).
\ee
\end{lemma}

Since
\bdis
\omega(t_0),m(t_0)\in (0,1),\ \Omega(t_0)>t_0\ln^4t_0,\ t_0\to\infty
\edis
(see (\ref{1.4}), (\ref{1.8}), then we obtain from (\ref{2.4}), (\ref{2.5}) by (\ref{1.6}) the following

\begin{lemma}
On the Riemann hypothesis we have
\be \label{2.6}
\begin{split}
& \sum_\gamma\frac{1}{(t-\gamma)^2},\ \left\{\frac{Z'(t)}{Z(t)}\right\}^2=\mcal{O}\left\{\frac{\ln^2t_0}{\omega^4(t_0)m^4(t_0)}\right\}, \\
& \alpha(t_0)\sum_\gamma\frac{1}{(t-\gamma)^2},\ \alpha^2(t_0)\left\{\frac{Z'(t)}{Z(t)}\right\}^2=\mcal{O}\left(\frac{1}{t_0}\right), \\
& t\in J(t_0),\ t_0\to\infty.
\end{split}
\ee
\end{lemma}

\section{Main equations of the relativistic cosmology and their incompleteness}

\subsection{}

Let us remind the Einstein's equations for the gravitation
\bdis
R^{\mu\nu}-\frac 12g^{\mu\nu}R+g^{\mu\nu}\Lambda=-\kappa c^2T^{\mu\nu},
\edis
where
\bdis
T^{\mu\nu}=\left( \rho+\frac{p}{c^2}\right)u^\mu u^\nu-g^{\mu\nu}\frac{p}{c^2},
\edis
is the energy-momentum tensor. In the case
\bdis
\begin{split}
& {\rm d}s^2={\rm d}t^2-\frac{R^2(t)}{c^2}\frac{{\rm d}r^2+r^2{\rm d}\theta^2+r^2\sin^2\theta{\rm d}\phi^2}{\left( 1+k\frac{r^2}{4}\right)^2},\ k=-1,0,1, \\
& u^1=u^2=u^3=0,\ u^4=1
\end{split}
\edis
we obtain the fundamental equations of the relativistic cosmology (comp. \cite{6}, p. 209)
\be \label{3.1}
\begin{split}
& \kappa c^2\rho=\frac{3}{R^2}(kc^2+R'^2)-\Lambda,\quad R'=\frac{{\rm d}R}{{\rm d}t}, \dots \\
& \kappa p=-\frac{2R''}{R}-\frac{R'^2}{R^2}-\frac{kc^2}{R^2}+\Lambda.
\end{split}
\ee
In (\ref{3.1}) we have: $R=R(t)$ is the \emph{radius of the Universe}, $\rho=\rho(t)$ and $p=p(t)$ denote the density and the pressure of the cosmic matter, and $\Lambda$ stands for the cosmological constant, $\kappa$ is the
Einstein's gravitational constant, and finally $c$ is the velocity of the light in the vacuum.

\begin{remark}
It is clear that the system of equations (\ref{3.1}) is incomplete. Namely, to make it complete we have to postulate some \emph{state equation}
\be \label{3.2}
G(\kappa c^2\rho,\kappa p)=0;\quad \kappa p=g(\kappa c^2\rho),
\ee
and after this we can solve the system of equations (\ref{3.1}).
\end{remark}

For example, when we postulate the state equation
\bdis
\kappa p=(\kappa c^2\rho)^{3/17};\quad \left(\frac{3}{17}\to a\in (0,1)\right)
\edis
then we obtain from (\ref{3.1}) the following agreeable differential equation for the function $R(t)$ (comp. (\ref{3.2})
\be \label{3.3}
\left(-\frac{2R''}{R}-\frac{R'^2}{R^2}-\frac{kc^2}{R^2}+\Lambda\right)^{17}=\left\{\frac{3}{R^2}(kc^2+R'^2)-\Lambda\right\}^3.
\ee

\subsection{}

For the purpose of this paper we will suppose in (\ref{3.1}) that $k=1$ (the spherical geometry) and $\Lambda>0$. Consequently, we will study the equations
\be \label{3.4}
\begin{split}
& kc^2\rho=3\left(\frac{R'}{R}\right)^2+3\frac{c^2}{R^2}-\Lambda, \\
& \kappa p=-\frac{2R''}{R}-\left(\frac{R'}{R}\right)^2-\frac{c^2}{R^2}+\Lambda.
\end{split}
\ee

\begin{remark}
In this paper:
\begin{itemize}
\item[(a)] we will postulate (instead of (\ref{3.2}) infinite set of the lines
\be \label{3.5}
R(t)=h\left(\left|\zf\right|\right),\quad t\to \infty ,
\ee
\item[(b)] after this we will define \emph{the physical domain} for the function $R(t)$,
\item[(c)] finally, we will study the set of the corresponding state equations
\bdis
\kappa p=g(\kappa c^2\rho),
\edis
(comp. \cite{7} and the papers \cite{10} -- \cite{12}).
\end{itemize}
\end{remark}

\begin{remark}
These are the reasons for our postulate (\ref{3.5}):
\begin{itemize}
\item[(a)] the aesthetic criterion based on the internal connection of the Riemann ideas with itself
\bdis
\text{Riemann}\left\{\begin{array}{lll} \text{Riemann's geometry} & \rightarrow & \text{Einstein's theory} \\
 & & \qquad \qquad \uparrow\\
 \text{Riemann's zeta-function} & \rightarrow & \text{Riemann's hypothesis} \end{array}\right.
\edis
(i. e. we wish to find some binding $\uparrow$).
\item[(b)] the almost random distribution of the members of the sequence $\{ \gamma''-\gamma'\}$, (comp., for example, the graph of the function $Z(t)$ in the neighborhood of the first Lehmer pair of the zeroes,
\cite{3}, p. 296),
\item[(c)] the Eddington discovery of the instability of the Einstein's spherical world.
\end{itemize}
\end{remark}

\subsection{}

In the case of the Einstein's universe (1917) -- the first cosmological application of the Einstein's theory of gravitation -- we have (comp. (\ref{3.2}))
\bdis
p(t)=0,\ R(t)=R_0,
\edis
and, from (\ref{3.4}) we obtain
\be \label{3.6}
R(t)=R_0=\frac{c}{\sqrt{\Lambda}},\ kc^2\rho(t)=2\Lambda.
\ee
Consequently, the Einstein's universe is described by the following triple
\be \label{3.7}
\{ R(t),\kappa c^2\rho(t),\kappa p(t)\}=\left\{\frac{c}{\sqrt{\Lambda}},2\Lambda,0\right\}.
\ee

\begin{remark}
Let us remind the instability of the Einstein's universe. This important fact was discovered by Eddington in 1930, (see \cite{2}, comp. \cite{1}, pp. 463-479).
Namely, from (\ref{3.4}), in the case $p(t)=0$ (see (\ref{3.7})) we obtain
\be \label{3.8}
6R''=(2\Lambda-kc^2\rho)R.
\ee
Consequently, if we have a small perturbation of the density $\rho(t)$ in (\ref{3.7}) such that
\bdis
\kappa c^2\rho<2\Lambda,
\edis
then the expansion (see (\ref{3.8})) of the universe follows and, if we have the small perturbation such that
\bdis
\kappa c^2\rho>2\Lambda,
\edis
then the contraction of the universe follows.
\end{remark}

\section{A new class of mathematical universes; some kindred of the Einstein's universe}

\subsection{}

In this paper we use the following postulate (see (\ref{1.11}), (\ref{3.5}), comp. \cite{10}, \cite{11})
\be \label{4.1}
\begin{split}
& R(t)=R(t;t_0,\Lambda,\mu)=\mu\frac{c}{\sqrt{\Lambda}}\left\{\frac{Z(t)}{Z(t_0)}\right\}^{\alpha(t_0)}=\\
&=\mu\frac{c}{\sqrt{\Lambda}}\left\{ 1+\mcal{O}\left(\frac{1}{t_0}\right)\right\},\quad t\in J(t_0),\ \mu>0.
\end{split}
\ee
Since by (\ref{3.8})
\be \label{4.2}
\frac{R'}{R}=\alpha(t_0)\frac{Z'(t)}{Z(t)},\ \frac{{\rm d}}{{\rm d}t}\left(\frac{R'(t)}{R(t)}\right)=\alpha(t_0)\frac{{\rm d}}{{\rm d}t}\left\{\frac{Z'(t)}{Z(t)}\right\},
\ee
then from (\ref{3.4}) by (\ref{1.11}), (\ref{2.6}), (\ref{4.1}), (\ref{4.2}) we have
\bdis
\begin{split}
& \kappa c^2\rho(t)=3\alpha^2(t_0)\left\{\frac{Z'(t)}{Z(t)}\right\}^2+\frac{3\Lambda}{\mu^2}\left\{\frac{Z(t_0)}{Z(t)}\right\}^{2\alpha(t_0)}-\Lambda=\\
& =\left(\frac{3}{\mu^2}-1\right)\Lambda +\mcal{O}\left\{\left( 1+\frac{1}{\mu^2}\right)\frac{1}{t_0}\right\}, \\
& \kappa p(t)=\alpha(t_0)\sum_\gamma\frac{1}{(t-\gamma)^2}-3\alpha^2(t_0)\left\{\frac{Z'(t)}{Z(t)}\right\}^2-\\
& - \frac{\Lambda}{\mu^2}\left\{\frac{Z(t_0)}{Z(t)}\right\}^{\alpha(t_0)}+\Lambda +\mcal{O}\left(\frac{1}{t_0}\right)=\\
&= \left(1-\frac{1}{\mu^2}\right)\Lambda+\mcal{O}\left\{\left(1+\frac{1}{\mu^2}\right)\frac{1}{t_0}\right\}.
\end{split}
\edis
Consequently, we have the following.

\begin{cor}
On the Riemann hypothesis we have the following infinite set of a mathematical universes
\be \label{4.3}
\begin{split}
& R(t;t_0,\Lambda,\mu)=\mu\frac{c}{\sqrt{\Lambda}}+\mcal{O}\left(\frac{1}{t_0}\right), \\
& \kappa c^2\rho(t;t_0,\Lambda,\mu)=\left(\frac{3}{\mu^2}-1\right)\Lambda+\mcal{O}\left\{\left(1+\frac{1}{\mu^2}\right)\frac{1}{t_0}\right\}, \\
& \kappa p(t;t_0,\Lambda,\mu)=\left( 1-\frac{1}{\mu^2}\right)\Lambda+\mcal{O}\left\{\left( 1+\frac{1}{\mu^2}\right)\frac{1}{t_0}\right\}, \\
& t\in J(t_0),\ \gamma'<t_0<\gamma'',\ \mu>0,\ t_0\to\infty.
\end{split}
\ee
\end{cor}

\subsection{}

The extremal state equations like these
\bdis
p(t)=c^2\rho(t),\quad p(t)=-c^2\rho(t)
\edis
are also used in the relativistic cosmology. We will define \emph{the physical domain} of the universe (\ref{4.3}) by means of the inequality (comp. \cite{10} -- \cite{12})
\be \label{4.4}
|p(t)|\leq c^2\rho(t).
\ee

\begin{mydef7}
Let
\be \label{4.5}
\begin{split}
& E_1(t;t_0,\Lambda,\mu)=\kappa c^2\rho-\kappa p, \\
& E_2(t;t_0,\Lambda,\mu)=\kappa c^2\rho+\kappa p.
\end{split}
\ee
Then we will call the set
\be \label{4.6}
F(t_0,\Lambda,\mu)=\{ t\in(\gamma',\gamma''):\ E_1(t)\geq 0,\ E_2(t)\geq 0,\rho(t)>0\},\ t_0\to\infty
\ee
\emph{the physical domain of the universe (\ref{4.3})}.
\end{mydef7}

Since for
\bdis
t\in J(t_0),\ t_0>K>0,
\edis
where $K$ is sufficiently big, we have (see (\ref{4.3}), (\ref{4.6}))
\bdis
\begin{split}
& E_1=2\left(\frac{2}{\mu^2}-1\right)\Lambda+\mcal{O}\left\{\left( 1+\frac{1}{\mu^2}\right)\frac{1}{t_0}\right\}, \\
& E_2=\frac{2}{\mu^2}\Lambda+\mcal{O}\left\{\left( 1+\frac{1}{\mu^2}\right)\frac{1}{t_0}\right\}, \\
& \kappa c^2\rho=\left(\frac{3}{\mu^2}-1\right)\Lambda+\mcal{O}\left\{\left( 1+\frac{1}{\mu^2}\right)\frac{1}{t_0}\right\},
\end{split}
\edis
and the inequalities are fulfilled for $\mu\in (0,\sqrt{2})$, then we have the following

\begin{cor}
On the Riemann hypothesis
\be \label{4.7}
J(t_0)\subset F(t_0,\Lambda,\mu),\quad \mu \in [\epsilon,\sqrt{2}-\epsilon],
\ee
where $\epsilon$ is an arbitrarily small fixed number.
\end{cor}

\begin{remark}
The essential part of the interval $(\gamma',\gamma'')$, $\gamma'<t_0<\gamma''$ belongs to the physical domain $F(t_0)$, (see (\ref{4.7}) and the Remark 3).
\end{remark}

\subsection{}

Next, we obtain from (\ref{4.3}), (\ref{4.7}) the following

\begin{cor}
On the Riemann hypothesis
\be \label{4.8}
\frac{p(t;t_0,\Lambda,\mu)}{c^2\rho(t;t_0,\Lambda,\mu)}\sim\frac{\mu^2-1}{3-\mu^2},\quad \mu\in [\epsilon,\sqrt{2}-\epsilon],\quad t\in J(t_0),\ t_0\to\infty.
\ee
\end{cor}

\begin{remark}
Consequently, on the Riemann hypothesis, we have obtained the continuum set of the asymptotically linear state equations (\ref{4.8}). For example:
\begin{itemize}
\item
\bdis
\mu=\epsilon \ \rightarrow \ p\sim -\frac 13(1-\epsilon')c^2\rho,
\edis

\item
\bdis
\mu=\frac{1}{\sqrt{2}} \ \rightarrow \ p\sim -\frac 13 c^2\rho,
\edis

\item
\bdis
\mu=1\ \rightarrow \ p\sim 0,
\edis
in this case we have the so-called incoherent dust (of galaxies), the small generalization of the Einstein's $p=0$,

\item
\bdis
\mu=\sqrt{\frac 65} \ \rightarrow \ p\sim \frac 19 c^2\rho,
\edis

\item
\bdis
\mu=\sqrt{\frac 32} \ \rightarrow \ p\sim \frac 13 c^2\rho,
\edis
in this case we have the universe filled by the photon gas,

\item
\bdis
\mu=\sqrt{2}-\epsilon \ \rightarrow \ p\sim (1-\epsilon'')c^2\rho,
\edis
\end{itemize}
where $0<\epsilon',\epsilon''$ are arbitrarily small values.
\end{remark}

\section{An infinite subset of microscopic universes  of the Einstein's type and the condition for the inflationary expansion of Universe}

\subsection{}

In the case
\bdis
\mu=\epsilon
\edis
with $\epsilon$ being an arbitrarily small fixed value, we obtain from (\ref{4.3}) the following infinite subset of the universes
\be \label{5.1}
\begin{split}
& R(t;t_0,\Lambda,\epsilon)=\epsilon\frac{c}{\sqrt{\Lambda}}+\mcal{O}\left(\frac{1}{t_0}\right), \\
& \kappa c^2\rho(t;t_0,\Lambda,\epsilon)=\left(\frac{3}{\epsilon^2}-1\right)\Lambda+\mcal{O}\left\{\left( 1+\frac{1}{\epsilon^2}\right)\frac{1}{t_0}\right\}, \\
& \kappa p(t;t_0,\Lambda,\epsilon)=\left( 1-\frac{1}{\epsilon^2}\right)\Lambda+\mcal{O}\left\{\left( 1+\frac{1}{\epsilon^2}\right)\frac{1}{t_0}\right\}, \\
& t\in J(t_0),\ \gamma'<t_0<\gamma'',\ t_0\to\infty.
\end{split}
\ee
We obtain for the volume of the universes (\ref{5.1})
\be \label{5.2}
V=V(t_0,\Lambda,\epsilon)=2\pi^2R^3=\frac{2\pi^2c^3}{\Lambda^{3/2}}\epsilon^3+\mcal{O}\left(\frac{1}{t_0}\right).
\ee
We can introduce also the local time $\tau$ for $J(t_0)$, namely
\bdis
\tau=\tau(t_0)=t-\{ t_0+\Delta_1(t_0)\},
\edis
where
\bdis
\tau\in [0,\Delta_1(t_0)+\Delta_2(t_0)],
\edis
and (see (\ref{1.4}), (\ref{1.7}))
\be \label{5.3}
\Delta_1(t_0)+\Delta_2(t_0)=\{ 1-\omega(t_0)\}(\gamma''-\gamma')<\gamma''-\gamma'<\frac{A}{\ln\ln\gamma'}\to 0
\ee
as $\gamma'\to \infty$. \\

Next, we have from (\ref{5.1}) that
\be \label{5.4}
\begin{split}
& \frac{p}{c^2\rho}=\frac
{\left(1-\frac{1}{\epsilon^2}\right)\Lambda+\mcal{O}\left\{\left( 1+\frac{1}{\epsilon^2}\right)\frac{1}{t_0}\right\}}
{\left(\frac{3}{\epsilon^2}-1\right)\Lambda+\mcal{O}\left\{\left( 1+\frac{1}{\epsilon^2}\right)\frac{1}{t_0}\right\}}=\\
& = \frac
{(\epsilon^2-1)\Lambda+\mcal{O}\left\{(1+\epsilon^2)\frac{1}{t_0}\right\}}
{(3-\epsilon^2)\Lambda+\mcal{O}\left\{(1+\epsilon^2)\frac{1}{t_0}\right\}}= \\
& = -\frac 13+\frac{2\epsilon^2}{9-3\epsilon^2}+\mcal{O}\left(\frac{1}{\Lambda t_0}\right).
\end{split}
\ee
Hence, we have the following

\begin{cor}
On the Riemann hypothesis there is an infinite set of the microscopic (see (\ref{5.2}), (\ref{5.3})) universes (\ref{5.1}) (a subset of the set
(\ref{4.3})) such that the state equation (see (\ref{5.4}))
\be \label{5.5}
\frac{p(t;t_0,\Lambda,\epsilon)}{c^2\rho(t;t_0,\Lambda,\epsilon)}=-\frac 13+\frac{2\epsilon^2}{9-3\epsilon^2}+\mcal{O}\left(\frac{1}{\Lambda t_0}\right),\
t\in J(t_0),\ t_0\to\infty.
\ee
\end{cor}

\begin{remark}
Let us remind that (see \cite{4}, p.13)
\bdis
L_P=8.10\times 10^{-35}cm
\edis
is the \emph{Planck length}, and
\bdis
T_P=2.70\times 10^{-43}s
\edis
is the \emph{Planck time}. In the case
\bdis
\epsilon\leq \frac{\sqrt{\Lambda}}{c}L_P,\quad \Delta_1(t_0)+\Delta_2(t_0)\leq T_P,
\edis
we have
\bdis
R(t;t_0,\Lambda,\epsilon)\leq L_P,\quad \tau\leq T_P,
\edis
i. e. we have in this case the infinite subset of the universes (\ref{5.1}) of the Planck scale.
\end{remark}

\subsection{}

Next, let us remind (see, for example, \cite{4}, p. 40) that the condition for the inflationary expansion of the universe can  be formulated
as follows
\be \label{5.6}
\text{INFLATION} \quad \Leftrightarrow \quad c^2\rho+3p<0
\ee
with
\bdis
\Lambda=0,
\edis
otherwise $\Lambda$ is absorbed into $c^2 \rho$ and $p$.

\begin{remark}
Consequently, the era of the inflation (that ends at $10^{-43}s$) is connected with the negative pressure $p$ because $\rho>0$ (always), and
\bdis
\frac{p}{c^2\rho}<-\frac 13.
\edis
\end{remark}

After this we can make the concluding remark.

\begin{remark}
The following holds true for the negative pressures:
\begin{itemize}
\item[(a)] if
\bdis
\frac{p}{c^2\rho}\in \left[\left. -1,-\frac 13\right)\right.
\edis
(for $-1$ see (\ref{4.4})) then we have the inflationary expansion of our universe,

\item[(b)] if (see (\ref{5.5}))
\bdis
\frac{p}{c^2\rho}=-\frac 13+\frac{2\epsilon^2}{9-3\epsilon^2}+\mcal{O}\left(\frac{1}{\Lambda t_0}\right)\in \left(-\frac 13,-\frac 13+\delta\right),
\edis
i. e. for a small right $\delta$-neighborhood of the point $-\frac 13$, we have, on the Riemann hypothesis, the infinite subset of the
microscopic universes of the Einstein's type.
\end{itemize}
\end{remark}

\section{Proof of Lemma 1}

Let
\be \label{6.1}
\sum_\gamma\frac{1}{(t-\gamma)^2}=\sum_{\gamma\leq \gamma'-1}+\sum_{\gamma\in (\gamma'-1,\gamma''+1)}+\sum_{\gamma''+1\leq \gamma}.
\ee
Since (see (\ref{1.4}))
\bdis
t-\gamma'\geq t_0-\Delta_1(t_0)-\gamma'= \omega(t_0)(t_0-\gamma'),
\edis
and, similarly,
\bdis
\gamma''-t\geq \omega(t_0)(\gamma''-t_0),
\edis
then (see (\ref{1.4}))
\be\label{6.2}
|t-\gamma|\geq \omega(t_0)m(t_0),\quad t\in J(t_0),\ \gamma\in (\gamma'-1,\gamma''-1).
\ee
Next, (comp. \cite{13}, p. 178)
\be \label{6.3}
\sum_{\gamma\in (\gamma'-1,\gamma''+1)}1=\mcal{O}(\ln t_0).
\ee
Then we obtain by (\ref{6.2}), (\ref{6.3})
\be \label{6.4}
\sum_{\gamma\in (\gamma'-1,\gamma''+1)}\frac{1}{(t-\gamma)^2}=
\mcal{O}\left\{\frac{\ln t_0}{\omega^2(t_0)m^2(t_0)}\right\},\quad t\in J(t_0).
\ee
Since
\bdis
|t-\gamma|=\gamma-t>\gamma-\gamma'',\quad t\in J(t_0),\ \gamma''+1\leq \gamma,
\edis
then (see (\ref{6.3}), comp. \cite{13}, p. 184)
\be \label{6.5}
\begin{split}
& \sum_{\gamma''+1\leq \gamma}\frac{1}{(t-\gamma)^2}<\sum_{\gamma''-1\leq \gamma}\frac{1}{(\gamma-\gamma'')^2}=\\
& =\sum_{n=1}^\infty\left\{\sum_{\gamma''+n\leq \gamma\leq \gamma''+n+1}\frac{1}{(\gamma-\gamma'')^2}\right\}< \\
& < A\sum_{n=1}^\infty \frac{\ln(\gamma''+n)}{n^2}<A\sum_{n\leq\gamma''}\frac{\ln 2\gamma''}{n^2}+\sum_{\gamma''<n}\frac{\ln 2n}{n^2}<\\
& < A\ln t_0,\quad t\in J(t_0).
\end{split}
\ee
By a similar way one can obtain the following estimate
\be \label{6.6}
\sum_{\gamma\leq \gamma'+1}\frac{1}{(t-\gamma)^2}< A\ln t_0,\quad t\in J(t_0).
\ee
Hence, we obtain (\ref{2.4}) from (\ref{6.1}) by (\ref{6.4}) -- (\ref{6.6}).

\section{Proof of Lemma 2}

First of all, we have by (\ref{2.4})
\bdis
\begin{split}
& \left|\sum_{\gamma}\frac{1}{(t-\gamma)(t_0-\gamma)}\right|\leq \left\{\sum_{\gamma}\frac{1}{(t-\gamma)^2}\right\}^{1/2}
\left\{\sum_{\gamma}\frac{1}{(t_0-\gamma)^2}\right\}^{1/2}<\\
& < A\frac{\ln t_0}{\omega^2(t_0)m^2(t_0)},\quad t\in J(t_0).
\end{split}
\edis
Next, by the Littlewood estimate (\ref{1.7}) we have
\bdis
|t-t_0|<\gamma''-\gamma'<\frac{A}{\ln\ln t_0},\quad t\in J(t_0).
\edis
Then, by (\ref{2.2}) we obtain the estimate
\bdis
\frac{Z'(t)}{Z(t)}=\mcal{O}\left\{\frac{\ln t_0}{\omega^2(t_0)m^2(t_0)\ln \ln t_0}\right\},\ t\in J(t_0),
\edis
and the estimate (\ref{2.5}) follows.

\section{Proof of Theorem}

Let
\be \label{8.1}
\begin{split}
& \sum_\gamma\left\{\frac{t-t_0}{t_0-\gamma}-\ln\left|\frac{t-\gamma}{t_0-\gamma}\right|\right\}=
\sum_{\gamma\leq \gamma'-1}+\sum_{\gamma''+1\leq \gamma}+\sum_{\gamma\in(\gamma'-1,\gamma''+1)}.
\end{split}
\ee

\subsection*{(A)}

If
\bdis
\gamma\leq \gamma'-1,
\edis
then we have for $t\in J(t_0)$ by (\ref{1.7}) that
\be \label{8.2}
\left|\frac{t-t_0}{t_0-\gamma}\right|\leq \frac{|t-t_0|}{t_0-\gamma'+1}<\gamma''-\gamma'<\frac{A}{\ln\ln t_0}.
\ee
Next,
\be \label{8.3}
\begin{split}
& \frac{t-t_0}{t_0-\gamma}-\ln\left|\frac{t-\gamma}{t_0-\gamma}\right|=\frac{t-t_0}{t_0-\gamma}-\ln\left( 1+\frac{t-t_0}{t_0-\gamma}\right)= \\
& = \left(\frac{t-t_0}{t_0-\gamma}\right)^2\sum_{k=0}^\infty\frac{(-1)^k}{k+2}\left(\frac{t-t_0}{t_0-\gamma}\right)^k,
\end{split}
\ee
\be \label{8.4}
\begin{split}
& \sum_{k=0}^\infty\frac{(-1)^k}{k+2}\left(\frac{t-t_0}{t_0-\gamma}\right)^k=\frac 12+
\mcal{O}\left\{\sum_{k=1}^\infty \left|\frac{t-t_0}{t_0-\gamma}\right|^k\right\}= \\
& =\frac 12+\mcal{O}\left(\frac{1}{\ln\ln t_0}\right).
\end{split}
\ee
Consequently,
\bdis
\begin{split}
& \sum_{\gamma\leq \gamma'-1}\left\{\frac{t-t_0}{t_0-\gamma}-\ln\left|\frac{t-\gamma}{t_0-\gamma}\right|\right\}=\\
& = (t-t_0)^2\sum_{\gamma\leq \gamma'-1}\frac{1}{(t_0-\gamma)^2}\left\{\frac 12+\mcal{O}\left(\frac{1}{\ln\ln t_0}\right)\right\},
\end{split}
\edis
and from this (see (\ref{1.7}), (\ref{6.6}))
\be \label{8.5}
0\leq \sum_{\gamma\leq \gamma'-1}\left\{\frac{t-t_0}{t_0-\gamma}-\ln\left|\frac{t-\gamma}{t_0-\gamma}\right|\right\}<
A\frac{\ln t_0}{(\ln\ln t_0)^2},\quad t\in J(t_0).
\ee

\subsection*{B} If
\bdis
\gamma''+1\leq \gamma ,
\edis
then we have for $t\in J(t_0)$, (comp. (\ref{8.2}))
\bdis
\left|\frac{t-t_0}{t_0-\gamma}\right|\leq \frac{|t-t_0|}{1+\gamma''-t_0}<\gamma''-\gamma'<\frac{A}{\ln\ln t_0},
\edis
and, similarly to (\ref{8.3}) -- (\ref{8.5}), we obtain
\be \label{8.6}
0\leq \sum_{\gamma''+1\leq \gamma}\left\{\frac{t-t_0}{t_0-\gamma}-\ln\left|\frac{t-\gamma}{t_0-\gamma}\right|\right\}<
A\frac{\ln t_0}{(\ln\ln t_0)^2},\quad t\in J(t_0).
\ee

\subsection*{C} Let
\bdis
\gamma\in (\gamma'-1,\gamma''+1),\quad t\in J(t_0)
\edis
and
\be \label{8.7}
\begin{split}
& V=\sum_{\gamma\in (\gamma'-1,\gamma''+1)}\left\{\frac{t-t_0}{t_0-\gamma}-\ln\left|\frac{t-\gamma}{t_0-\gamma}\right|\right\}= \\
& =\sum_{\gamma\in (\gamma'-1,\gamma''+1)}\frac{t-t_0}{t_0-\gamma}+\sum_{\gamma\in (\gamma'-1,\gamma''+1)}\ln\left|\frac{t_0-\gamma}{t-\gamma}\right|
=V_1+V_2.
\end{split}
\ee
Since
\bdis
\left|\frac{t-t_0}{t_0-\gamma}\right|\leq \frac{Q(t_0)}{m(t_0)},\quad t\in J(t_0),\ \gamma\in (\gamma'-1,\gamma''+1),
\edis
then (see (\ref{1.2}), (\ref{6.3}), (\ref{8.7}))
\be \label{8.8}
|V_1|\leq \sum_{\gamma\in (\gamma'-1,\gamma''+1)}\left|\frac{t-t_0}{t_0-\gamma}\right|<
A\ln t_0\cdot t_0\ln^3t_0=At_0\ln^4t_0.
\ee
Next, by (\ref{1.2}), (\ref{6.2}) we have
\be \label{8.9}
\begin{split}
& \left|\frac{t_0-\gamma}{t-\gamma}\right|=\left| 1+\frac{t_0-t}{t-\gamma}\right|\leq 1+\frac{|t-t_0|}{|t-\gamma|}<1+\frac{Q(t_0)}{\omega(t_0)m(t_0)}< \\
& < \frac{2}{\omega(t_0)}\frac{Q(t_0)}{m(t_0)}<\frac{2}{\omega(t_0)}t_0\ln^3 t_0.
\end{split}
\ee
Since (see (\ref{1.4}), (\ref{1.7}))
\be \label{8.10}
\begin{split}
& |t_0-\gamma|\geq m(t_0),\quad \gamma\in (\gamma'-1,\gamma''+1), \\
& |t-\gamma|=\gamma-t<\gamma''+1-\gamma'=\gamma''-\gamma'+1<2,\quad \gamma\in [\gamma'',\gamma''+1), \\
& |t-\gamma|=t-\gamma<\gamma''-\gamma'+1<2,\quad \gamma\in (\gamma'-1,\gamma'],
\end{split}
\ee
then (see (\ref{8.9}), (\ref{8.10}))
\bdis
\frac{1}{2}m(t_0)<\left|\frac{t_0-\gamma}{t-\gamma}\right|<\frac{2}{\omega(t_0)}t_0\ln^3 t_0.
\edis
Consequently,
\bdis
-\ln\frac{2}{m(t_0)}<\ln\left|\frac{t_0-\gamma}{t-\gamma}\right|<\ln\left\{\frac{2}{\omega(t_0)}t_0\ln^3t_0\right\},
\edis
(of course, $m(t_0)\in (0,1),\ t_0\to\infty$), and (see (\ref{1.4}))
\be \label{8.11}
\left|\ln\left|\frac{t_0-\gamma}{t-\gamma}\right|\right|<\ln w(t_0),\quad t\in J(t_0),\ \gamma\in (\gamma'-1,\gamma''+1).
\ee
Next, (see (\ref{6.3}), (\ref{8.7}), (\ref{8.11}))
\be \label{8.12}
|V_2|\leq \sum_{\gamma\in (\gamma'-1,\gamma''+1)}\left|\ln\left|\frac{t_0-\gamma}{t-\gamma}\right|\right|<A\ln t_0\ln w(t_0).
\ee
Next, (see (\ref{8.7}), (\ref{8.8}), (\ref{8.12}))
\be \label{8.13}
|V_1+V_2|\leq |V_1|+|V_2|<At_0\ln^4t_0+A\ln t_0\ln w(t_0),
\ee
and, of course,
\be \label{8.14}
V_1+V_2>-At_0\ln^4t_0-A\ln t_0\ln w(t_0).
\ee
Finally, the estimate (\ref{1.5}) follows from (\ref{2.3}) by (\ref{8.1}), (\ref{8.5}), (\ref{8.6}), (\ref{8.13}) and (\ref{8.14}).

\thanks{I would like to thank Michal Demetrian for helping me with the electronic version of this work.}

\end{document}